\documentclass[12pt]{article}

\newcommand{\be}{\begin{equation}}
\newcommand{\ee}{\end{equation}}
\newcommand{\bi}[1]{\vspace{-3mm} \bibitem{#1}}

\usepackage{epsfig,amsmath,amssymb,graphics,graphicx} 

\begin{document}
\begin{center}

{\it Journal of Mathematical Physics. 
Vol.55. No.8. (2014) 083510.}
\vskip 3mm

{\bf \large Anisotropic Fractal Media by \\
\vskip 3mm
Vector Calculus in Non-Integer Dimensional Space } 

\vskip 7mm
{\bf \large Vasily E. Tarasov} \\
\vskip 3mm

{\it Skobeltsyn Institute of Nuclear Physics,\\ 
Lomonosov Moscow State University, Moscow 119991, Russia} \\
{E-mail: tarasov@theory.sinp.msu.ru} \\

\begin{abstract}
A review of different approaches to describe
anisotropic fractal media is proposed.
In this paper differentiation and integration 
non-integer dimensional  and multi-fractional spaces 
are considered as tools to describe
anisotropic fractal materials and media.
We suggest a generalization of vector calculus for
non-integer dimensional space by using 
a product measure method. 
The product of fractional and non-integer dimensional spaces
allows us to take into account the anisotropy of the fractal media in the framework of continuum models.
The integration over non-integer-dimensional spaces 
is considered.
In this paper differential operators of first and second orders
for fractional space and non-integer dimensional space 
are suggested.
The differential operators are defined as inverse
operations to integration in spaces with non-integer dimensions.
Non-integer dimensional space that is product
of spaces with different dimensions
allows us to give continuum models for anisotropic 
type of the media.
The Poisson's equation for fractal medium, 
the Euler-Bernoulli fractal beam, and
the Timoshenko beam equations for fractal material
are considered as examples of application of
suggested generalization of vector calculus for
anisotropic fractal materials and media.
\end{abstract}

\end{center}


\section{Introduction}


Fractals are measurable metric sets with non-integer 
dimensions \cite{Fractal1,Fractal2}. 
The basic property of the fractal is non-integer 
Hausdorff dimension that should be observed at all scales. 
The definition of the Hausdorff dimension 
requires the diameter of the covering sets to vanish.
The fractal structure of real materials 
cannot be observed on all scales.
This structure exists only for scales $R >R_0$, 
where $R_0$ is the characteristic size of 
atoms or molecules of fractal media.
Isotropic fractal materials can be characterized by 
the relation between
the mass $M_D(W_B)$ of a ball region $W_B$ of fractal medium,
and the radius $R$ of this ball in the form
\be \label{MDW-1}
M_D(W_B) =M_0 \ \left(\frac{R}{R_0}\right)^D , 
\quad R/R_0 \gg 1 . 
\ee
The parameter $D$ is called the 
mass dimension of fractal medium. 
The parameter $D$, does not depend on the shape 
of the region $W_B$, or on whether the packing of sphere 
of radius $R_0$ is close packing, a random packing 
or a porous packing with a uniform distribution of holes.
Anisotropic fractal materials can be characterized by 
the power-law relation for the mass of
the parallelepiped region $W_P$ in the form
\be \label{MDW-2}
M_D(W_P) = M_0 
\left(\frac{L_x}{R_0}\right)^{\alpha_1} \, 
\left(\frac{L_y}{R_0}\right)^{\alpha_2} \, 
\left(\frac{L_z}{R_0}\right)^{\alpha_3} , \quad
\operatorname{min} \{L_x,L_y,L_z\} \gg R_0 , 
\ee
where the parameter $\alpha_k$ is non-integer dimension 
along $X_k$-axis, $k=1,2,3$,
and $L_x$, $L_y$, $L_z$ represent three edges
that meet at one vertex.
The parameter $\alpha_k$ describes how to increase 
the medium mass in the case of increasing the size of 
the parallelepiped along one axis,
when the  parallelepiped sizes along other axes do not change.
The sum $D=\alpha_1+\alpha_2+\alpha_3$ is called 
the fractal mass dimension of the anisotropic fractal medium.

Using (\ref{MDW-1}) and (\ref{MDW-2}),  
we can define a fractal material as a medium 
with non-integer mass dimension. 
The non-integer dimension does not reflect 
completely all specific properties of the fractal media,
but it is an main characteristic of fractal media and materials.
For this reason, we assume that continuum models 
with non-integer dimensional spaces allow us 
to derive important conclusions about 
the behavior of the fractal media.

The main ways of describing of fractal media can be 
conventionally divided into the following five approaches.

1) {\it Analysis on fractal approach}:
The first approach is based on the use of
methods of "Analysis on fractals" 
\cite{Kugami,Strichartz-1,Strichartz-2,Harrison,Kumagai,DGV}. 
Unfortunately a possibility of application of the 
"Analysis on fractals" to solve differential equations 
on fractals \cite{Strichartz-1} for real problems 
of fractal materials 
is very limited due to weak development of 
this area of mathematics to the present time.

2) {\it Fractional-differential continuum model approach}:
The second approach is based on the use of
the fractional derivatives of non-integer orders
with respect to space coordinates
to describe some properties of fractal materials \cite{CM}.
Therefore the correspondent models 
can be called the fractional-differential models.
It has been suggested 
by Carpinteri and co-workers in \cite{CCC2001,CCC2002,CCC2003},
where so-called local fractional derivatives are used, 
and then developed in \cite{CCC2004a,CCC2004b,CCC2004c,CCC2004d,CCC2004e,CCSPZ2009,CCC2009}.
Unfortunately there are not enough differential equations 
with these fractional derivatives that are solved 
for various problems of fractal materials.
It should be noted that the usual Leibniz rule 
does not hold \cite{CNSNS2013}
for derivatives of non-integer orders (and integer orders 
$n \ne 1$). It is a characteristic property of fractional derivatives.

3) {\it Fractional-integral continuum model approach}:
The third approach has been suggested in 
\cite{PLA2005-1,AP2005-2,PLA2005-2,IJMPB2005-2,MPLB2005-1,CMDA2006,TarasovSpringer} and it is based on application of
continuum models of fractal media. 
These models can be called fractional-integral continuum 
model because the integrations of non-integer orders are used.
The kernels of fractional integrals are defined by 
power-law density of states \cite{TarasovSpringer}.
The orders of fractional integrals are equal to the
mass (charge or other physical) dimensions of media.
In these models, the density of states is applied in addition
to the notion of distribution functions such as 
density of mass, density of charge.
There are a lot of applications of these continuum models
in different fields of mechanics and physics
(see \cite{TarasovSpringer} and references cited therein).
These models have been applied by Ostoja-Starzewski in
\cite{MOS-01,MOS-02,MOS-03,MOS-04,MOS-05}.
A generalization of fractional-integral continuum models 
for anisotropic fractal media
has been suggested by Ostoja-Starzewski and co-workers 
in \cite{MOS-3,MOS-4,MOS-4b,MOS-5,MOS-6,MOS-6b,MOS-7}.
In these models, the differential operators
are modified by the density of states.
However, these operators have integer-order 
differential operators.

4) {\it Fractional space approach:}
The fourth approach uses the concept of
a fractional space, which is characterized by 
non-integer (fractional) powers of coordinates.
The fractional space approach has been suggested in papers 
\cite{Chaos2004,PRE2005,JPCS2005} 
and then it is used for applications in different areas
\cite{Chaos2005,IJMPB2006,IJMPB2007,MPLB2007,TarasovSpringer}
(see also \cite{POP2005,POP2006,MPLA2006,MPLB-2005}).
This approach has been developed by Calcagni in  
\cite{Calcagni1,Calcagni2,Calcagni3,Calcagni4}
(see also \cite{Calcagni-A1,Calcagni-A2}),
and then it was generalized for anisotropic case
by using a multi-fractional space ("multi-scale space")
in \cite{Calcagni-B1,Calcagni-B2,Calcagni-B3}.
The first interpretation of the fractional phase space 
is connected with fractional dimension space. 
The fractional dimension interpretation follows from
the formulas for dimensional regularizations
and it was suggested in \cite{Chaos2004}.
In the paper \cite{PRE2005,JPCS2005} the second interpretation 
of the fractional phase space is considered. 
This interpretation follows from 
the fractional measure \cite{Chaos2004} of phase space 
that is used in the fractional integrals, i.e.
the integrals of non-integer orders.
In the third interpretation, the fractional phase space 
is considered as a phase space that is described 
by the fractional powers of coordinates and momenta.
In addition almost all Hamiltonian systems 
with fractional phase space are non-Hamiltonian 
dissipative systems in the usual phase space. 
It allows us to have the fourth interpretation of 
the fractional phase space
as a phase space of power-law type of non-Hamiltonian systems.
Using fractional space approach we
can consider wide class of non-Hamiltonian systems 
as generalized Hamiltonian systems. 
Differentiation in fractional space approach
can be used in two forms: 
(a) the usual derivatives with respect to fractional powers of coordinates \cite{Chaos2004,PRE2005,JPCS2005}; 
(b) the fractional derivatives of non-integer orders (fractional derivatives) with respect to coordinates
\cite{Calcagni1,Calcagni2,Calcagni3,Calcagni4}.
The term "fractional space" is sometimes used
for non-integer-dimensional space.
This leads to confusion and misunderstanding.
We use the term "fractional space"
for effective space $\mathbb{R}^n$ with coordinates that
are non-integer powers of coordinates of physical space.
In the fractional space approach, 
the integer-dimensional spaces $\mathbb{R}^n$,
the integration and differentiation of integer-orders
for these spaces are used.
The coordinates of fractional space are considered 
as effective coordinates that are fractional powers
of real space coordinates of physical system.
Note that we also can use effective spaces
for non-integer-dimensional physical spaces.
Expressions for the effective coordinates 
for fractional and non-integer-dimensional spaces
differ by factors in the density of states.

5) {\it Non-integer-dimensional space approach}: 
The fifth approach is based on application of
integration and differentiation 
for non-integer-dimensional spaces. 
The integration in non-integer dimensional space
is well developed \cite{Wilson,Stillinger,Collins}, and
it has a wide application in quantum field theory.
The axioms for integrals in $D$-dimensional space 
is suggested in \cite{Wilson}. 
This properties are natural and 
necessary in applications \cite{Collins}.
Integration in $D$-dimensional spaces 
with non-integer $D$ is used for dimensional regularization 
in quantum field theory \cite{HV1972,Leibbrandt,Collins}
and in physical kinetics \cite{WilsonFisher,WK1974}.
Dimensional regularization is a way to get infinities 
that occur when one evaluates 
Feynman diagrams in quantum theory. 
Differentiation in non-integer dimensional space is
considered in \cite{Stillinger,PS2004,CNSNS2014}. 
In the papers \cite{Stillinger,PS2004} 
it was offered only a scalar Laplacian for 
non-integer dimensional space.
Unfortunately the gradient, divergence, curl operator 
and the vector Laplacian \cite{VLap}
are not considered in \cite{Stillinger,PS2004}. 
The scalar Laplace operators, whcih are suggested 
by Stillinger in \cite{Stillinger} 
and Palmer, Stavrinou in \cite{PS2004}
for non-integer dimensional spaces, 
have successfully been used for effective descriptions 
in different areas of physics and mechanics
such as quantum mechanics 
(see \cite{Stillinger,PS2004}, 
\cite{XFHe1} - \cite{XFHe4},
\cite{Thilagam1997a} - \cite{Thilagam2005},
\cite{MA1997} - \cite{MA2003c},
\cite{Muslih2009} - \cite{SPL2013},
the diffusion processes  \cite{LSTLRL}, 
the general relativity \cite{SMB,SM},
and the electrodynamics \cite{MB2007,BGG2010,MSBR2010,ZMN2010,ZMN2011a,ZMN2011b,ZMN2011c}. 
All these applications are based only on 
two generalization of the scalar Laplacian 
that are suggested in \cite{Stillinger,PS2004}. 
To expand the range of possible applications
of models with non-integer dimensional spaces
it is important to have generalization of
differential operators of first orders 
(gradient, divergence, curl operators) 
and the vector Laplacian. 
The continuation in dimension 
is recently suggested in \cite{CNSNS2014,CSF2014} to define
the gradient, divergence, curl operator and the vector Laplacian 
for non-integer dimensional space. 
It allows us to describe isotropic fractal media
in the framework of continuum models
with non-integer dimensional spaces.
To generalize non-integer dimensional space approach
for anisotropic fractal media we can suggest to use
the product measure approach  
suggested in \cite{Chaos2004,PRE2005,JPCS2005} and \cite{PS2004}.
Generalizations of the gradient, divergence, 
curl operators and the vector Laplace operator 
for non-integer dimensional and fractional spaces 
to describe anisotropic fractal media
are not considered by the product measure 
approach in \cite{Stillinger,PS2004} and other papers.
These generalizations are suggested in this paper
as an extension of approach proposed in \cite{CNSNS2014,CSF2014}. 

To present more clearly some of the differences between 
these five approaches to describe fractal media
distributed in the space $\mathbb{R}^n$, 
we present the following table. \\

\noindent
\begin{tabular}{|p{0.23\linewidth}|p{0.22\linewidth}| p{0.2\linewidth}| p{0.22\linewidth}|}
\hline
Approach & Set / Space & Integration & Differentiation \\
\hline
Analysis on fractal &  
Fractal set & Integration for $ \quad $ fractal set &
Differentiation for $ \quad $   fractal set \\
\hline
Fractional-differential continuum  model &  
Integer-dimensional space $\mathbb{R}^n$&  
Integrals for $\mathbb{R}^n$ &  
Fractional-order derivatives for $\mathbb{R}^n$ \\
\hline
Fractional-integral continuum model & 
Integer-dimensional space $\mathbb{R}^n$&  
Fractional-order integrals for $\mathbb{R}^n$ &  
Integer-order \, derivatives for $\mathbb{R}^n$\\
\hline
Fractional space &
Integer-dimensional (effective) space $\mathbb{R}^n$&  
Integer-order \quad \quad integrals for $\mathbb{R}^n$ &  
Integer-order \, derivatives for $\mathbb{R}^n$\\
\hline
Non-integer-dimensional space &  
Non-integer-dimensional space &  
Integrals for $ \quad $ non-integer-dimensional space &  
Derivatives for non-integer-dimensional space \\
\hline
\end{tabular} 
\\
\\

Let us note some advantages and disadvantages of the
the third, fourth and fifth approaches.
The forms of the functions that define the density of states 
in the third and fourth approaches have arbitrariness due 
to the existence of different types of fractional integrals.
In the fifth approach, the form of density of states 
is uniquely fixed by the expression of volume
of the region in the non-integer-dimensional space.
In the third approach practically all properties of 
fractal media are reduced to coordinate transformations
and to space curvature of power type.
The transition by transformations to an effective Euclidean space
cannot be considered as a consistent approach
to describe fractal media.
These transformations can be considered only as a part
of mathematical method to solve some equations
in the framework of continuum models
with non-integer-dimensional space, i.e., 
in the  fifth approach.
Connections between the effective coordinates
and physical coordinates
for the non-integer-dimensional space approach
are uniquely defined.
In addition, the fifth approach allows us to use
fractional derivatives and integrals to describe
nonlocal type of fractal media that cannot be described
by other approaches.

In this paper, we consider the two last approaches 
based on the fractional space and
the non-integer dimensional space
to describe anisotropic fractal materials and media.
We suggest a generalization of vector calculus
for non-integer dimensional space that is product
of spaces with different dimensions.
In section 2, we discuss the product measure method
to describe anisotropic fractal media.
In section 3, the integration for fractional 
and non-integer-dimensional spaces are considered.
In section 4, differential operations of first and second orders
for fractional space and non-integer-dimensional space 
are suggested.
In section 5, we give some examples of application of
suggested generalization of vector calculus for
anisotropic fractal materials and media.


\section{Product measure method}

\subsection{Product measure for fractional 
and non-integer-dimensional spaces }

The product measure method can be applied to
the non-integer dimensional spaces and to the fractional spaces:

(I) {\it Product measure for the fractional spaces}.
The product measure approach for the fractional spaces
has been suggested in \cite{PRE2005,JPCS2005},
where fractional phase space is considered with its interpretation as a non-integer (fractional) dimensional space.
In papers \cite{PRE2005,JPCS2005}, the following measure is used for generalized coordinates and momenta
\be
d\mu_{\alpha} (x_k) = c(\alpha) \, |x_k|^{\alpha-1} \, dx_k ,
\ee
where the numerical factor $c(\alpha)$ is
\be \label{caG}
c(\alpha) = 1/\Gamma(\alpha).
\ee 
We use the factor $c(\alpha)$ in the form (\ref{caG}) to get 
a relation with the Riemann-Liouville fractional integrations
of non-integer order $\alpha$.
In the papers \cite{Chaos2004,PRE2005,JPCS2005}, 
it has been shown that the integration
for the suggested fractional space
is directly connected with integration 
for non-integer dimensional space up to numerical factor.
Therefore the fractional space has been interpreted 
as a non-integer dimensional space.
The suggested fractional space approach has been used in 
\cite{Chaos2005,IJMPB2006,IJMPB2007,MPLB2007,TarasovSpringer} 
for configuration and phase spaces,
and then it has been applied by Calcagni in 
\cite{Calcagni1,Calcagni2,Calcagni3,Calcagni4}
for space-time.
The differentiation and integration
in fractional space are considered as
differentiation and integration with respect to
non-integer (fractional) powers of coordinates.
The differential operator of the first order is defined 
\cite{Chaos2004,PRE2005,JPCS2005,Chaos2005,IJMPB2006,IJMPB2007} 
by
\be
D_{\alpha,k} = \frac{\partial}{\partial Q_k} =
\frac{1}{\alpha \, |x_k|^{\alpha-1}} 
\frac{\partial}{\partial x_k} ,
\ee
where $Q_k= \operatorname{sgn}(x) \, |x|^{\alpha}$.
The product measure approach also used 
in \cite{MOS-3,MOS-4,MOS-4b,MOS-5,MOS-7} 
to describe fractal media,
but the fractional space as space with non-integer powers of coordinates has not been considered.
Instead of the Riemann-Liouville fractional integration,
which is used in \cite{Chaos2004,PRE2005,JPCS2005,Chaos2005,IJMPB2006,IJMPB2007,MPLB2007}, 
the product measure approach is used 
in \cite{MOS-3,MOS-4,MOS-4b,MOS-5,MOS-7} 
for so-called modified Riemann-Liouville integrations 
in the integer dimensional spaces. \\


(II)  {\it Product measure for the non-integer dimensional spaces}. 
The product measure approach has been suggested
by Palmer and Stavrinou in \cite{PS2004}
for non-integer dimensional spaces,
where each orthogonal coordinates has own dimension. 
The methods of this paper can be considered as
a modification of Stillinger's \cite{Stillinger}
and Svozils's \cite{Svozil} methods 
for the product measure approach.
In the paper \cite{PS2004}, the product of the following 
single-variable measures is used 
\be \label{PSm-1}
d\mu_{\alpha_k} (x_k) = 
c(\alpha_k) \, |x_k|^{\alpha_k -1} \, dx_k  
\ee
with the numerical factor
\be \label{PSm-1b}
c(\alpha) = \frac{2 \, \pi^{\alpha/2}}{\Gamma(\alpha/2)} .
\ee
Note that $c(\alpha)$ is equal to 
the surface area $S_{\alpha}(R)$ of $\alpha$-sphere 
with the radius $R=1$, where
\be
S_{\alpha}(R) =  \frac{2 \, \pi^{\alpha/2}}{\Gamma(\alpha/2)} \, R^{\alpha} .
\ee
The integration for non-integer dimensional space 
by product measure approach is described in \cite{PS2004}.
The scalar Laplacian operator for non-integer dimensional spaces 
is also suggested in \cite{PS2004} in the form
\be \label{PS-LAP}
^S\Delta^{(\alpha)}=
\sum^3_{k=1} \Bigl( \frac{\partial^2}{\partial x^2_k}
+\frac{\alpha_k-1}{x_k}  \frac{\partial}{\partial x_k}\Bigr) ,
\ee
where $(\alpha)=(\alpha_1,\alpha_2,\alpha_3)$
is the multi-index. Unfortunately definitions 
of the gradient, divergence, curl operations, and 
the vector Laplacian are not considered in \cite{PS2004}. \\


\subsection{Type of anisotropic fractal media 
described by non-integer dimensional spaces}

In this section, we describe a type of anisotropic fractal media 
that can be considered by approach
based on non-integer dimensional spaces.

Let us defined the parameter $\alpha_k$ ($k=1;2;3$)
that describes the scaling property along $X_k$-axis by
\be \label{alpha-1}
\alpha_1 = \alpha_x = D - d_{yz} ,
\ee
where $D$ is a non-integer mass dimension of the material.
Here $d_{yz}$ is the non-integer dimension 
(for example, the box-counting dimension) 
of the $YZ$-cross-section, 
which is perpendicular to the $X$-axis. 
We should assume that non-integer dimension $d_{yz}$
of the $YZ$-cross-sections 
is the same for all points along the $X$-axis
($d_{yz}(x)=d_{yz}=\operatorname{const}$).
If this condition is not satisfied, then we have 
a variable order $\alpha_1(x)$ that depends on the coordinates.

In general, the parameters $d_{xy}$, $d_{xz}$, $d_{yz}$
cannot be considered as dimensions of the boundaries
of fractal media.
The boundary of fractal medium can be fractal surface
with dimension $D_s$. 
In the general case, $D_s$ can be greater than two.
The non-integer dimensions 
of the cross-section are always less than two,  
\be
0<d_{xy} <2, \quad 0<d_{xz} <2, \quad 0<d_{yz} <2 .
\ee
Similarly (\ref{alpha-1}), we can define other two parameters 
$\alpha_2=\alpha_y$ and $\alpha_3=\alpha_z$.

Let ${\bf L}_1$, ${\bf L}_2$, and ${\bf L}_3$ 
be the basis vectors that define a three-dimensional 
parallelepiped in $\mathbb{R}^3$.
The parallelepiped region is the convex hull for these vectors
\be \label{W-P}
W_P:=\left\{ \sum^{n=3}_{k=1} a_k \, {\bf L}_k: \quad 0 \le a_k \le 1 \right\} .
\ee
For simplification we will consider the rectangular parallelepiped only.
Anisotropic fractal material can be characterized by 
the power-law relation for the mass $M_D(W_P)$ of
the parallelepiped region (\ref{W-P}) in the form
\be
M_D(W_P) = M_0 
\left(\frac{L_x}{R_0}\right)^{\alpha_1} \, 
\left(\frac{L_y}{R_0}\right)^{\alpha_2} \, 
\left(\frac{L_z}{R_0}\right)^{\alpha_3} ,
\ee
where the parameter $\alpha_k$ is non-integer dimension 
along $X_k$-axis, $k=1,2,3$, and
$L_k=|{\bf L}_k|$ are the magnitudes of 
vectors ${\bf L}_k$, $k=1,2,3$.
The values $L_k$ can be considered as three edges
that meet at one vertex, and
$R_0$ is a characteristic size of particles
(atoms or molecules) of fractal medium.
The parameter $\alpha_k$ describes how to increase the mass medium in the case of increasing the size of 
the parallelepiped region along one axis,
when the  parallelepiped sizes along other axes do not change.
The sum $D=\alpha_1+\alpha_2+\alpha_3$ is called 
the dimension of the anisotropic fractal medium.

In general, the parameters $\alpha_k>0$ ($k=1,2,3$) 
can be either less than unity or greater than unity.
In all cases, the following conditions should be satisfied
\be
0<\alpha_1+\alpha_2+\alpha_3 = D \le 3 .
\ee
For example, the conditions $0<\alpha_k<1$ 
for all $k \in \{1,2,3\}$ 
hold for fractal media similar to 3D Cantor dust.
We assume that the parameter $\alpha_k >1$ 
describes a fractal flow and motion of medium in $X_k$-direction.
This possibility is based on an assumption 
that trajectories of the medium particles 
in the $X_k$-direction are fractal curve
with the dimension $\alpha_k>1$.
For example, the Koch curve has 
$\alpha_k=\ln(4) / \ln(3) \approx 1.262$.
We also assume that $\alpha_k>1$ can be used for materials 
consisting of fractal molecular curves or fractal chains 
\cite{TarasovSpringer}.


\subsection{Single-variable measure}

For the fractional space approach and
the non-integer dimensional space approach,
we can use the single-variable measure  
\be
d \mu (\alpha, x) = c(\alpha) \, |x|^{\alpha-1} \, dx ,
\ee
where $\alpha>0$ is a parameter that will be considered 
as a non-integer (fractional) dimension of the line, 
and $c(\alpha)$ is a function of $\alpha$. 
Here, we take the absolute value of $x$ in $|x|^{\alpha-1}$ 
to consider positive and negative values of $x$.
For $\alpha=1$, the numerical factor $c(\alpha)$ 
must be equal to $1$ in order to have
\[ d \mu(1,x) = dx . \]
Using the product measure approach for $\mathbb{R}^3$   
with point coordinates $x_1$, $x_2$, $x_3$, 
the single-variable measures are  
\be
d \mu (\alpha_k,x_{k}) = 
c(\alpha_k) \, |x_k|^{\alpha_k-1} \, dx_k .
\ee 
In the product measure method we can use the following 
two ways to define a numerical factor $c(\alpha_k)$.
In the first way of description, 
the factor is defined by a connection with
integrals of non-integer orders $\alpha_k$
in the integer dimensional space.
In the first way, the factor is defined by a connection 
with integrals in spaces with non-integer 
dimensions $\alpha_k$ along the $X_k$-axis.

Let us give the effective coordinates for these two cases.

(1) The fractional space approach is based on the use 
of the following new (effective) coordinates 
\be \label{Q-k}
Q_k= Q_k(\alpha_k,x_k) = \frac{1}{\Gamma (\alpha_k+1)} \, 
\operatorname{sgn}(x_k) \, |x_k|^{\alpha} 
\ee
that is connected with the single-variable measure of the form
\be
d \mu (\alpha_k,x_k) = d Q_k = 
\frac{1}{\Gamma (\alpha_k)} \, |x_k|^{\alpha_k-1} \, dx_k ,
\ee
where the numerical factor in the density of states is
\be
c(\alpha) = \frac{1}{\Gamma(\alpha)} .
\ee
This form of $c(\alpha_k)$ is based on the connection
of the Riemann-Liouville integrals of 
non-integer orders $\alpha_k$.

(2) For non-integer dimensional space approach, 
we can use the effective coordinates
\be \label{X-k}
X_k = X_k(\alpha_k,x_k) = \frac{\pi^{\alpha_k/2}}{2 \, \Gamma (\alpha_k/2+1)} \, 
\operatorname{sgn}(x_k) \, |x_k|^{\alpha_k} ,
\ee
that is connected with the single-variable measure \cite{Stillinger,PS2004} of the form
\be \label{mu-k}
d \mu (\alpha_k, x_k) = d X_k = 
\frac{\pi^{\alpha_k/2}}{\Gamma (\alpha_k/2)} 
\, |x_k|^{\alpha_k-1} \, dx_k .
\ee
Here we take the density of states in the form
\be
c(\alpha) = \frac{\pi^{\alpha/2}}{\Gamma(\alpha/2)} ,
\ee
such that the area of the sphere with radius $r$ is equal to
\be \label{SD-1}
S_{\alpha-1}(r)= 
\frac{2 \, \pi^{\alpha/2}}{\Gamma(\alpha/2)} \, r^{\alpha-1} .
\ee
The absolute values $|x_k|$ can be interpreted 
as radii $r_k=|x_k|$ 
of sphere with non-integer dimension $\alpha_k$.
The presence of a factor of $2$ for $S_{\alpha-1}$ 
in (\ref{SD-1}) is due to the fact that 
for $\alpha=1$, the variable $r$ is integrated from $-R$ to $R$, 
and when the limits are taken as $0$ and $R$, 
one gets a factor of $2$. 

The space with coordinates (\ref{X-k}) and 
the product of single-variable measures (\ref{mu-k})
can be considered as a non-integer dimensional space. 
The parameter $D = \alpha_1 + \alpha_2 + \alpha_3$ 
can be interpreted as a dimension of the space. 
For $\alpha_1 = \alpha_2 = \alpha_3=1$, we get $D=3$, i.e., 
the dimension of space is the usual integer dimension. 
If $\alpha_1 = \alpha_2 = \alpha_3=\alpha$, 
where $0< \alpha \le 1$, we have a non-integer dimensional space 
for isotropic fractal materials.
Regardless of the isotropic or the anisotropic case, 
the dimension of the space is given by 
\[ D = \alpha_1 + \alpha_2 + \alpha_3 . \] 
We have a non-integer dimensional space 
if at least one of the parameters $\alpha_k$ is not equal to $1$.

\subsection{Density of states}

A connection between the single-variable measure 
$d\mu (\alpha_k,x_k)$ of non-integer dimensional space
and the measure $d\mu (1,x_k)$ of integer dimensional space is
\be
d\mu (\alpha_k,x_k) = c_1(\alpha_k,x_k) \, d\mu (1,x_k) 
\ee
The functions $c_1(\alpha_k,x_k)$ should 
be considered as a density of states
of fractal material \cite{TarasovSpringer}
along the $X_k$-axis.
We can define the density of states along 
the $X_k$-axis by the equation
\be c_1(\alpha_x,x) = 
\frac{c_3(D, x,y,z)}{c_2(d_{yz},y,z)} , \ee
where $c_3(D, x,y,z)$ is the density of state 
in the volume of material,
and $c_2(d_{xy},x,y)$, $c_2(d_{xz},x,z)$,
$c_2(d_{yz},y,z)$  are 
density of states of the $XY$, $XZ$, $YZ$- 
cross-sections respectively.

The interpretation of the functions
$c_3(D, x,y,z)$,$c_2(d_{xy},x,y)$, $c_2(d_{xz},x,z)$,
$c_2(d_{yz},y,z)$, and $c_1(\alpha_k,x_k)$, $(k=1,2,3)$, 
as densities of states
has been suggested in \cite{TarasovSpringer}.
The density of states $c_n$ describes how closely 
packed permitted states of particles in the space $\mathbb{R}^n$.
In the book \cite{TarasovSpringer} 
the density of states is defined by 
integrations of non-integer orders
that is also called the fractional integration.
For $x_k \in [a_k;b_k]$ the function $c_1(\alpha_k,x_k)$ 
is defined in \cite{TarasovSpringer} by
\be
c_1(\alpha_k,x_k) = 
\frac{1}{\Gamma (\alpha_k)} \, |x_k-a_k|^{\alpha_k-1} .
\ee
This form is connected with 
the Riemann-Liouville fractional integral.
In the paper \cite{MOS-3}, an expression 
for $c_1(\alpha_k,x_k)$ has been suggested in the form
\be
c_1(\alpha_k,x_k) = \alpha_k \, |b_k-x_k|^{\alpha_k-1} 
\ee
that does not contain the gamma function.
This form is connected with 
the modified Riemann-Liouville fractional integral.

In this paper, we use integration 
in non-integer dimensional space 
instead of fractional integration.
Then we should use the density of states in the form
\be \label{c1-1}
c_1(\alpha_k,x_k) = 
\frac{\pi^{\alpha_k/2}}{\Gamma (\alpha_k/2)} \, 
|x_k|^{\alpha_k-1} 
\ee
that is defined by the measure for integration 
in non-integer dimensional space.
An application of density of states in the form (\ref{c1-1})
allows us to get the expression for lengths 
\be
\int^R_{-R} d \mu (\alpha,x)= 
2 \int^R_0 d \mu (\alpha,x)= 
\frac{2 \pi^{\alpha/2}}{\alpha \Gamma (\alpha/2)} \, R^{\alpha} =
\frac{\pi^{\alpha/2}}{\Gamma (\alpha/2+1)} \, R^{\alpha} =
V_{\alpha}(R) 
\ee
that coincides with the well-known value 
for non-integer dimensional volume.

\section{Integration in fractional and non-integer dimensional spaces}

\subsection{Product spaces and product measures}

The integral for non-integer dimensional space
is defined for a single-variable  in \cite{Svozil}.
It is useful for integrating spherically 
symmetric functions only.
We can consider multiple variables by using the product spaces 
and product measures \cite{PS2004}.

Using a collection of $n=3$
measurable sets $(W_k,\mu_k,D)$ with $k=1,2,3$,
we form a Cartesian product $W=W_1 \times W_2 \times W_3$
of the sets $W_k$.
The definition of product measures and 
an application of Fubini's theorem gives 
a measure for the product set $W=W_1 \times W_2 \times W_3$ as
\be
\mu_B(W)=
(\mu_{\alpha_1} \times \mu_{\alpha_2} \times \mu_{\alpha_3})(W)
= \prod^{n=3}_{k=1} \mu (\alpha_k,W_k) .
\ee
Then integration over a function $f$ on $W$ is
\be \label{1-int-n}
\int_W f (x_1,x_2, x_3) \, d\mu_B
=\int_{W_1} \int_{W_2} \int_{W_3}
f (x_1,x_2 , x_3) \, \prod^{n=3}_{k=1} \, d \mu (\alpha_k,x_k) .
\ee
In this form, the single-variable measure 
may be used for each coordinate $x_k$, 
which has an associated non-integer dimension $\alpha_k$, 
by the equation
\be
d \mu (\alpha_k,x_k) = c_1(\alpha_k,x_k) \, dx_k , 
\quad (k=1,2,3),
\ee
where $c_1(\alpha_k,x_k)$ is the density of states
of the form
\be \label{DOS-NIDS}
c_1(\alpha_k,x_k) = \frac{\pi^{\alpha_k/2}}{\Gamma(\alpha_k/2)}
\; |x_k|^{\alpha_k-1} .
\ee
Note that we use $c_1(\alpha_k,x_k)$ without the factor 2.

Then the total dimension of $W=W_1 \times W_2 \times W_3$ is
$D = \alpha_1+\alpha_2+\alpha_3 $.

\subsection{Reproduce the single-variable integration}

Let us reproduce the result for 
the single-variable integration in the form
\be \label{1-intWf}
\int_W f (x_1,x_2, x_3) \, d\mu_B
= \frac{2 \pi^{D/2}}{\Gamma(D/2)}
\int^{\infty}_0 f(r) \, r^{D-1} \, dr 
\ee
for spherically symmetric function $f (x_1, x_2 , x_3) = f (r)$
in $W_1 \times W_2 \times W_3$,
where $r^2 = (x_1)^2 + (x_2)^2 + (x_3)^2$. 
For this function, we can perform 
the integration in spherical coordinates $(r, \phi, \theta)$.
The Cartesian coordinates $(x_1,x_2,x_3)$ 
can be expressed by the spherical coordinates 
($r$, $\varphi$, $\theta$), where 
$r \in [0, \infty)$, $\varphi \in [0, 2\pi)$, 
$\theta \in [0,\pi]$, by:
\be
x_1=r \, \sin\theta \, \cos\varphi ,
\ee
\be
x_2=r \, \sin\theta \, \sin\varphi ,
\ee
\be
x_3=r \, \cos\theta .
\ee

In this case, equation (\ref{1-int-n}) becomes
\[
\int_W d \mu_B \; f (x_1,x_2,x_3) =
\]
\[ =A(\alpha_1,\alpha_2,\alpha_3)
 \int_{W_1} d x_1 \int_{W_2} d x_2 \int_{W_3}
d x_3 \, |x_1|^{\alpha_1-1} |x_2|^{\alpha_2-1} |x_3|^{\alpha_3-1} \, f(x_1,x_2,x_3)=  \]
\[
=A(\alpha_1,\alpha_2,\alpha_3) \int^{\infty}_0 dr
\int^{2\pi}_{0} d \varphi \int^{\pi}_{0} d \theta 
\, J(r,\theta) \; r^{ \alpha_1+\alpha_2+ \alpha_3- 3} \, 
\cdot \]
\[ \cdot \,
|\cos \varphi|^{\alpha_1 - 1} \, 
|\sin \varphi|^{\alpha_2 - 1}
|\sin \theta|^{\alpha_1+\alpha_2 - 1} 
|\cos \theta|^{\alpha_3 - 1} 
\, f(r) ,
\]
where
\be
A(\alpha_1,\alpha_2,\alpha_3)=
\frac{\pi^{\alpha_1/2}}{\Gamma(\alpha_1/2)}
\frac{\pi^{\alpha_2/2}}{\Gamma(\alpha_2/2)}
\frac{\pi^{\alpha_3/2}}{\Gamma(\alpha_3/2)} ,
\ee
and $J(r,\phi)= r^2 \, \sin \theta$
is the Jacobian of the coordinate change.

Since the function is only dependent on the radial variable and
not the angular variables, we get the product of three integrals
\[
\int_W d \mu_B \; f (x_1,x_2,x_3) =
A(\alpha_1,\alpha_2,\alpha_3) \,
\int^{\infty}_0 \, f(r) \, r^{ \alpha_1+\alpha_2+ \alpha_3- 1} 
\, \cdot
\]
\be
\cdot \,
\int^{2\pi}_{0} d \varphi \,
|\cos \varphi|^{\alpha_1 - 1} \, |\sin \varphi|^{\alpha_2 - 1}
\int^{\pi}_{0} d \theta \,
|\sin \theta|^{\alpha_1+\alpha_2 - 1} \,
|\cos \theta|^{\alpha_3 - 1} .
\ee
Using equation (26) from Section 2.5.12 of \cite{PBM-1},
in the form
\be
\int^{\pi/2}_0 (\sin x)^{\mu-1} \, (\cos x)^{\nu-1} \, dx =
\frac{\Gamma(\mu/2) \, \Gamma(\nu/2)}{2 \, \Gamma((\mu+\nu)/2)} ,
\ee
where $\mu >0$, $\nu>0$, we have
\[
\int^{2\pi}_{0} d \varphi \,
|\cos \varphi|^{\alpha_1 - 1} \, |\sin \varphi|^{\alpha_2 - 1}=
\]
\be
=4 \, \int^{\pi/4}_{0} d \varphi \,
(\cos \varphi)^{\alpha_1 - 1} \, (\sin \varphi)^{\alpha_2 - 1}=
\frac{2 \, \Gamma(\alpha_1/2) \,\Gamma(\alpha_2/2)}{\Gamma((\alpha_1+\alpha_2)/2)} ,
\ee
\[
\int^{\pi}_{0} d \theta \,
|\sin \theta|^{\alpha_1+\alpha_2 - 1} \,
|\cos \theta|^{\alpha_3 - 1} = 
\]
\be
=2\, \int^{\pi/2}_{0} d \theta \,
(\sin \theta)^{\alpha_1+\alpha_2 - 1} \,
(\cos \theta)^{\alpha_3 - 1} =
\frac{\Gamma((\alpha_1+\alpha_2)/2) \,\Gamma(\alpha_3/2)}{\Gamma((\alpha_1+\alpha_2+\alpha_3)/2)} .
\ee
Using $D=\alpha_1+\alpha_2+\alpha_3$, we obtain for 
$f (x_1,x_2, x_3)=f(r)$ the ralation
\be \label{import}
\int_W f (x_1,x_2, x_3) \,
d\mu_1(x_1) d \mu_2(x_2) d\mu_3(x_3)  =
\frac{2 \, \pi^{D/2}}{\Gamma(D/2)} 
\int^{\infty}_0 f(r) \, r^{D-1} \, dr .
\ee
This equation describes the $D$-dimensional integration
of a spherically symmetric function,
and reproduces the result (\ref{1-intWf}).

It is important to note that relation (\ref{import}) holds
only for density of states $c(\alpha_k,x_k)$
that corresponds to 
the non-integer dimensional space (\ref{DOS-NIDS}).
Equation (\ref{import}) cannot hold
only for density of states $c_1(\alpha_k,x_k)$
suggested in \cite{AP2005-2} and \cite{MOS-3}.


\section{Differential operators for fractional and non-integer dimensional spaces}


\subsection{Laplace operator for non-integer dimensional and fractional spaces}

Let us consider possible definitions of the 
scalar Laplace operators for non-integer dimensional and fractional spaces.

(I) {\it Scalar Laplace operators for non-integer dimensional space}.
The Laplacian operator for non-integer dimensional space
has been suggested in \cite{PS2004} in the form
\be \label{PS-LAP2}
^S\Delta^{(\alpha)}=
\sum^3_{k=1} \Bigl( \frac{\partial^2}{\partial x^2_k}
+\frac{\alpha_k-1}{x_k}  \frac{\partial}{\partial x_k}\Bigr) ,
\ee
where $(\alpha)=(\alpha_1,\alpha_2,\alpha_3)$. 
Unfortunately definitions of differential operators
of first order such as gradient, divergence, 
curl operations are not considered in \cite{PS2004}. 
The Laplacian operator (\ref{PS-LAP2}) 
is used in \cite{ZMN2010,ZMN2011a,ZMN2011b,ZMN2011c,ZMN}.
In the book \cite{ZMN}, 
the gradient, divergence, and curl operators
are suggested only as approximations of 
the square of the Laplace operator (\ref{PS-LAP}). 
The first order differential operator proposed in \cite{ZMN}
is considered as an approximation of (\ref{PS-LAP2}) in the form
\be \label{ZMN-D1}
D_{\alpha_k,k} \approx 
\frac{\partial}{\partial x_k} + 
\frac{\alpha_k-1}{2\, x_k} .
\ee
For example, the gradient is 
\be \label{ZMN-D2}
\operatorname{grad}_D \varphi \approx \sum^3_{k=1}
\Bigl( \frac{\partial \varphi}{\partial x_k} + 
\frac{\alpha_k-1}{2\, x_k} \varphi \Bigr) \, {\bf e}_k ,
\ee
and the divergence for vector field ${\bf u}=u_k \, {\bf e}_k$ 
is defined by
\be \label{ZMN-D3}
\operatorname{div}_D {\bf u} \approx \sum^3_{k=1}
\Bigl( \frac{\partial u_k}{\partial x_k} + 
\frac{\alpha_k-1}{2\, x_k} u_k \Bigr) .
\ee
Obviously, the corresponding scalar Laplacian 
\be \label{ZMN-Lap}
\operatorname{div}_D \operatorname{grad}_D \varphi \approx
\sum^3_{k=1} \left(
\frac{\partial^2 \varphi}{\partial x^2_k} + 
\frac{\alpha_k-1}{x_k} \frac{\partial \varphi}{\partial x_k} + \frac{(\alpha_k-1)(\alpha_x-3)}{4\, x^2_k} \varphi \right)
\ee
does not coincide with the operator (\ref{PS-LAP2}).


(II) {\it Scalar Laplace operators for fractional space}.
The scalar Laplace operators 
have been suggested by Calcagni \cite{Calcagni4} 
for for fractional space-time.
We can represent the suggested equations for fractional space
in the following forms
\be \label{K1a}
{\cal K}_1 \varphi = 
\sum^3_{k=1}  \frac{1}{v(\alpha,x)} 
\frac{\partial}{\partial x_k} \left( v(\alpha,x) \,
\frac{\partial \varphi}{\partial x_k} \right) ,
\ee
\be \label{K2a}
{\cal K}_2 \varphi = 
\sum^3_{k=1}  \frac{1}{\sqrt{v(\alpha,x)}} 
\frac{\partial^2}{\partial x^2_k} 
\Bigl( \sqrt{v(\alpha,x)} \, \varphi \Bigr) ,
\ee
and
\be \label{K3a}
{\cal K}_{\alpha,l} \varphi = 
\sum^3_{k=1}  \frac{ (x_k)^{l-1/2} }{\sqrt{v(\alpha,x)}} 
\frac{\partial}{\partial x_k} \left( (x_k)^{l-1/2}
\frac{\partial}{\partial x_k} 
\Bigl( (x_k)^{l-1/2} \, \sqrt{v(\alpha,x)} \, \varphi \Bigr) \right) , 
\ee
where $x_k > 0$, and $v(\alpha,x)$ is 
the isotropic measure weight
\be \label{vK}
v(\alpha,x) = \prod^3_{k=1} c_1(\alpha,x_k) =
\prod^3_{k=1} \frac{(x_k)^{\alpha-1}}{\Gamma(\alpha)} ,
\quad (x_k \ge 0 ) .
\ee
Here $c_1(\alpha,x_k)$ is the density of states 
for fractional space \cite{TarasovSpringer}.
We should note that Calcagni consider the Laplace operators
for Minkowski space-time $\mathbb{R}^4_{1,3}$.
Equations (\ref{K1a}), (\ref{K2a}) and (\ref{K3a})  
are given for the Euclidean space $\mathbb{R}^3$
to describe fractal material.
Note that the expression (\ref{K3a}) contains 
(\ref{K1a}) and (\ref{K2a}) as particular cases.
The operators (\ref{K1a}), (\ref{K2a}) can be represented as
\be
{\cal K}_1 = {\cal K}_{\alpha,l=1-\alpha/2}  , \quad
{\cal K}_2 = {\cal K}_{\alpha,l=1/2} .
\ee
Substitution of (\ref{vK}) into (\ref{K1a}), 
(\ref{K2a}), (\ref{K3a}) gives 
the Laplace operators in the forms
\be \label{K1b}
{\cal K}_1 \varphi = \sum^3_{k=1} \left( 
\frac{\partial^2 \varphi}{\partial x^2_k} + 
\frac{\alpha-1}{x_k} \frac{\partial \varphi}{\partial x_k} \right) ,
\ee
\be \label{K2b}
{\cal K}_2 \varphi = \sum^3_{k=1} \left( 
\frac{\partial^2 \varphi}{\partial x^2_k} + 
\frac{\alpha-1}{x_k} \frac{\partial \varphi}{\partial x_k} + \frac{(\alpha-1)(\alpha-3)}{4\, x^2_k} \varphi \right) ,
\ee
\be \label{K3b}
{\cal K}_{\alpha,l} \varphi = 
\sum^3_{k=1} \left( 
\frac{\partial^2 \varphi}{\partial x^2_k} + 
\frac{\alpha-1}{x_k} \frac{\partial \varphi}{\partial x_k} + \frac{(\alpha-2)^2 -4 \, l^2}{4\, x^2_k} \varphi  \right) .
\ee
The Laplace operator (\ref{K1b}) 
coincides with the expression (\ref{PS-LAP}) 
suggested in \cite{PS2004}.
The Laplace operator (\ref{K2b}) 
coincides with the operator (\ref{ZMN-Lap}) 
that can be derived from the first order 
operators (\ref{ZMN-D2}) and (\ref{ZMN-D2})
suggested in \cite{ZMN}.
It was proved \cite{Calcagni4} that
the case $l=1/2$ is unique because it is only one, 
where the Laplace operator of the type (\ref{K3a}) 
can be represented
as the square of first order differential operator.
This first order operator 
(see equation (3.43) in \cite{Calcagni4}) is
\be \label{Del1}
{\cal D}_{\alpha,k} \varphi = 
\frac{1}{\sqrt{v(\alpha,x)}} 
\frac{\partial}{\partial x_k} 
\Bigl( \sqrt{v(\alpha,x)} \, \varphi \Bigr) ,
\ee
such that 
\be
{\cal K}_2 \varphi = \sum^3_{k=1} 
\Bigl( {\cal D}_{\alpha,k} \Bigr)^2 \varphi .
\ee
Substitution of (\ref{vK}) into (\ref{Del1}) gives 
\be \label{Del2}
{\cal D}_{\alpha,k} \varphi = 
\frac{\partial\varphi }{\partial x_k} 
+ \frac{\alpha-1}{2\, x_k} \, \varphi .
\ee
We can see that operator (\ref{Del2})
can be considered as (\ref{ZMN-D1}) 
with $\alpha_k=\alpha$ for all $k=1,2,3$.

To anisotropic fractal media and materials 
in the framework of the non-integer dimensional space approach
we should generalize the isotropic 
measure weight $v(\alpha,x)$ by using 
\be \label{vK2}
v(\alpha,x) = \prod^3_{k=1}
\frac{\pi^{\alpha_k/2}}{\Gamma(\alpha_k/2+1)}  
|x_k|^{\alpha_k-1} 
\ee
instead of the expression (\ref{vK}) for fractional space.
The derivative of the first order (\ref{Del2})
should also be generalized for anisotropic fractal media as
\be \label{Del2-an}
{\cal D}_{\alpha,k} \varphi = 
\frac{\partial\varphi }{\partial x_k} 
+ \frac{\alpha_k-1}{2\, x_k} \, \varphi ,
\ee
where we take into account different values of dimensions
along the $X_k$-axis.
Using the first order differential operator (\ref{Del2-an}),
it is easy to define the del operator, gradient, divergence,
curl operators, and the vector Laplacian 
in order to describe fractal media
in the framework of the non-integer dimensional space approach.

\subsection{Approaches to formulation of 
the vector calculus in non-integer dimensional space}


In the Stillinger's paper \cite{Stillinger}, 
the scalar Laplace operator 
for non-integer dimensional space is suggested only.
Generalizations of gradient, divergence, curl operators, 
and the vector Laplacian are not considered in \cite{Stillinger}.
A generalization of the gradient, divergence, curl operators,
the scalar and vector Laplace operators for 
non-integer dimensional space can be defined
by the method of continuation in dimension 
\cite{CNSNS2014,CSF2014}.
For simplification, only radial case
is consider in \cite{CNSNS2014}, where the scalar and vector 
fields are independent of the angles, and 
the vector fields are directed along the radius vector.
This simplification is analogous to consideration 
of the integration in non-integer dimensional space 
in Section 4 of \cite{Collins}.  


The main advantage of the product measure approach
to define the vector calculus for
non-integer dimensional space
is a possibility to describe anisotropic fractal materials.
In this paper, we suggest differential vector operators for
non-integer dimensional space by product measure method 
to describe anisotropic fractal media 
in the framework of continuum models.
The differential operators are defined as an inverse
operation to integration in non-integer dimensional space.

We can state that it is possible to reduce 
non-integer dimensional space
to a power-law curved space by considering 
the density of states $c_1(\alpha_k,x_k)$ 
as the Lame coefficients 
\be
H_k = c_1(\alpha_k,x_k) .
\ee

We consider Euclidean space for 
the effective coordinates $X_k(x_k)$,
which are defined by (\ref{X-k}), such that
\be \label{d2sX}
d^2 s_X = \sum^3_{k=1} (dX_k)^2 = 
\sum^3_{k=1} c^2_1(\alpha_k,x_k) \, (dx_k)^2, 
\ee
where $c_1(\alpha_k,x_k)$ are the density of states of the form 
\be \label{c-1-H}
c_1(\alpha_k,x_k) = \frac{\pi^{\alpha_k/2}}{\Gamma (\alpha_k/2)} \, |x_k|^{\alpha_k-1} .
\ee
Using (\ref{d2sX}), we can see that  
the densities of states (\ref{c-1-H}) are the Lame coefficients 
\be
H_k =\sqrt{ \sum^{n=3}_{k=1} 
\left( \frac{\partial X_i}{\partial x_k} \right)^2 } .
\ee
It allows us to use the following well-known equations and definitions.
The metric tensors of the Euclidean space 
in curvilinear coordinates $x^k$ are
\be \label{metric-t}
g_{kl}(x) = H^2_k \, \delta_{kl} , \quad
g^{kl}(x) = \frac{1}{H^2_k} \, \delta_{kl} ,
\ee
and indices can be raised and lowered by this metric
\be
u^k = g^{kl}(x) \, u_l , \quad u_k = g_{kl} \, u^l .
\ee
For an orthogonal basis, we have
\be
g= \operatorname{det}(g_{kl}) = \prod^{n=3}_{k=1} g_{kk} = 
\prod^{n=3}_{k=1} H^2_k , 
\ee
and
\be
J =\sqrt{|g(x)|} = \prod^{n=3}_{k=1} H_k =H_1 \, H_2 \, H_3 .
\ee
Then the volume differential 3-form is given by
\be
\operatorname{vol}_{\alpha} = 
\sqrt{|g(x)|} \ dx^1 \wedge dx^2 \wedge dx^3 .
\ee
The correspondent volume for arbitrary set $W$ in local coordinates is
\be
\operatorname{vol}_{\alpha} (W) = 
\int_W \sqrt{|g(x)|} \, dx^1 \, dx^2 \, dx^3 .
\ee

Using the metric tensor (\ref{metric-t}),
the gradient of scalar function $\varphi$ is the vector field 
\be
\operatorname{grad}_{(\alpha)} \varphi = \sum^{n=3}_{k,l=1}
{\bf e}_k \, g^{kl}(x) \, \frac{\partial \varphi}{\partial x_l} ,
\ee 
and the divergence of vector field ${\bf u}= u^k \, {\bf e}_k$ 
is defined as the scalar function by
\be
\operatorname{div}_{(\alpha)} {\bf u} = \sum^{n=3}_{k=1}
\frac{1}{\sqrt{|g(x)|}} 
\frac{\partial (\sqrt{|g(x)|} \, u_k) }{\partial x^k} .
\ee
The Laplace-Beltrami operator 
is defined as the divergence of the gradient.
Using the definitions of the gradient and divergence, 
the Laplace-Beltrami operator applied 
to a scalar function $\varphi$ is given 
in local coordinates by
\be
\Delta_{(\alpha)} \varphi =
\operatorname{div}_{(\alpha)} 
\operatorname{grad}_{(\alpha)} \varphi = \sum^{n=3}_{k=1}
\frac{1}{\sqrt{|g|}} \frac{\partial}{\partial x^k} 
\left( \sqrt{|g(x)|} \, g^{kl}(x) \frac{\partial \varphi}{\partial x^l}\right) .
\ee

Using the effective coordinates $X_k= X_k(\alpha_k,x_k)$,
which are defined in (\ref{X-k}), 
we can define the nabla operator (the del operator) by
\be
\nabla_{(\alpha)} = \sum^{3}_{k=1} 
{\bf e}_k \, \frac{\partial}{\partial X_k} =  \sum^{3}_{k=1}  
{\bf e}_k \, \frac{1}{c_1(\alpha_k,x_k)} \, 
\frac{\partial}{\partial x_k} ,
\ee
where $(\alpha)=(\alpha_1,\alpha_2,\alpha_3)$ is the multi-index,
and $c_1(\alpha_k,x_k)$ is defined by (\ref{c-1-H}). 
We can use the well-known relations for the gradient, 
divergence, the curl operator, 
the scalar and vector Laplace operators
through the Lame coefficients.
The gradient for scalar field
\be \label{grad-alpha}
\operatorname{grad}_{(\alpha)} \varphi = \sum^3_{k=1}
\frac{1}{H^2_k} \frac{\partial \varphi}{\partial x_k} 
\, {\bf e}_k .
\ee
The divergence for vector field
\be \label{div-alpha}
\operatorname{div}_{(\alpha)} {\bf u} = \sum^3_{k=1}
\frac{1}{H_1 \, H_2 \, H_3} \frac{\partial}{\partial x_k} 
\left( \frac{H_1 \, H_2 \, H_3}{H_k} u_k\right) .
\ee
The curl operator for vector field
\be \label{curl-alpha}
\operatorname{curl}_{(\alpha)} {\bf u} = 
\sum^3_{k,i,j=1} \frac{1}{H_1 \, H_2 \, H_3} \,
{\bf e}_i \epsilon_{ijk} \, H_i
\frac{\partial (H_k \, u_k)}{\partial x_j} ,
\ee
where $\epsilon_{ijk}$ is the Levi-Civita symbol.
The scalar Laplacian
\be
\Delta_{(\alpha)} \varphi = \sum^3_{k=1}
\frac{1}{H_1 \, H_2 \, H_3} \frac{\partial}{\partial x_k} 
\left( \frac{H_1 \, H_2 \, H_3}{H^2_k} 
\frac{\partial \varphi }{\partial x_k} \right) .
\ee

Using that $H_k = c_1(\alpha_k,x_k)$,
we get $\partial H_k/\partial x_l=0$ for $k \ne l$, and
the divergence 
\be
\operatorname{div}_{(\alpha)} {\bf u} = \sum^3_{k=1}
\frac{1}{H_k} \frac{\partial u_k}{\partial x_k} .
\ee
The curl operator for vector field
\be
\operatorname{curl}_{(\alpha)} {\bf u} = 
\sum^3_{k,i,j=1} \frac{1}{H_j} \,
{\bf e}_i \epsilon_{ijk} 
\frac{\partial u_k}{\partial x_j} ,
\ee
where $\epsilon_{ijk}$ is the Levi-Civita symbol.
The scalar Laplacian
\be
\Delta_{(\alpha)} \varphi = \sum^3_{k=1}
\frac{1}{H_k} \frac{\partial}{\partial x_k} 
\left( \frac{1}{H_k} \,
\frac{\partial \varphi}{\partial x_k} \right) .
\ee


We can define a differential operator that takes into account 
the density of states $c_1(\alpha_k,x_k)$ by
\be \label{partial-k}
\partial_{x_k, \alpha_k} = \frac{\partial}{\partial X_k}=
\frac{1}{c_1(\alpha_k,x_k)} \frac{\partial}{\partial x_k} ,
\ee
where $c_1(\alpha_k,x_k)$ corresponds to 
the non-integer dimensionality along the $X_k$-axis
and it is defined by (\ref{c-1-H}).  
These derivatives cannot be considered as derivatives of non-integer orders
(as fractional derivatives) or as fractal derivatives 
(derivatives on fractal set).
The operators $\partial_{x_k, \alpha_k}$ 
are usual differential operators of the first order that 
is defined for differentiable functions on $\mathbb{R}^3$.

The form of derivatives (\ref{partial-k}) is analogous to 
differential operator suggested in \cite{MOS-7}.
The main difference is that we define the operators 
(\ref{partial-k}) for the non-integer dimensional spaces 
in the framework of the measure product approach that is
described by Palmer and Stavrinou \cite{PS2004} 
(see also \cite{MPLA2006,POP2006,TarasovSpringer}). 
The differential operators suggested in \cite{MOS-7} are defined 
for modified Riemann-Liouville fractional 
integral of orders $\alpha_k$
in the integer dimensional space.

Using the operators (\ref{partial-k}), we can introduce 
generalized vector differential operations.
The gradient
\be \label{grad-a1}
\operatorname{grad}_{(\alpha)} \varphi({\bf r}) = \sum^3_{k=1}
{\bf e}_k \, \partial_{x_k, \alpha_k} \varphi({\bf r}) ,
\ee
where ${\bf e}_k$ are unit base vector of the Cartesian coordinate system.
The divergence of the vector field ${\bf u} ({\bf r}) ={\bf e}_k \, u_k ({\bf r})$  is
\be \label{div-a1}
\operatorname{div}_{(\alpha)} {\bf u}({\bf r}) = 
\sum^3_{k=1} \partial_{x_k, \alpha_k} \, u_k ({\bf r}) .
\ee
The curl for the vector field ${\bf u} ({\bf r}) = {\bf e}_k \, u_k ({\bf r})$  is
\be
\operatorname{curl}_{(\alpha)} {\bf u}({\bf r}) =  \sum^3_{k,i,l=1} 
{\bf e}_i \, \epsilon_{ikl} \, \partial_{x_k, \alpha_k} \, u_l ({\bf r}) ,
\ee
where $\epsilon_{ikl}$ is the Levi-Civita symbol (or alternating symbol).

Using (\ref{grad-a1}) and (\ref{div-a1}) we can define the
second order differential operators such as the scalar Laplacian
and vector Laplacian.
The scalar Laplacian has the from
\be \label{SLap-alpha}
^S\Delta_{(\alpha)} \varphi ({\bf r}) =
\operatorname{div}_{(\alpha)} \operatorname{grad}_{(\alpha)} 
\varphi ({\bf r}) .
\ee
The vector Laplacian \cite{VLap} has the from
\be \label{VLap-alpha}
^V\Delta_{(\alpha)} {\bf u} ({\bf r}) =
\operatorname{grad}_{(\alpha)} \operatorname{div}_{(\alpha)} 
{\bf u}({\bf r}) -
\operatorname{curl}_{(\alpha)} \operatorname{curl}_{(\alpha)} 
{\bf u}({\bf r}) .
\ee
The scalar Laplacian (\ref{SLap-alpha}) can be represented 
by using the usual partial derivatives
\be \label{NEW-S}
^S\Delta_{(\alpha)} \varphi ({\bf r}) =
\sum^3_{k=1} 
\frac{1}{c^2_1(\alpha_k,x_k)}
\left( \frac{\partial^2 \varphi}{\partial x^2_k} 
- \frac{\alpha_k-1}{x_k} \frac{\partial \varphi}{\partial x_k} 
\right) , 
\ee
where $c_1(\alpha_k,x_k)$ is  density of states (\ref{c-1-H})
along the $X_k$-axis for model with
the non-integer dimensional space.  
It is easy to see that this operator does not coincide 
with the Laplace operators proposed 
by Palmer and Stavrinou \cite{PS2004}, 
by Stillinger \cite{Stillinger}, and 
by Calcagni \cite{Calcagni4}. 
The main advantage of the suggested Laplace 
operator (\ref{NEW-S}) is that firstly, 
it is obtained as the action of the gradient and divergence, 
and secondly, it is adapted for models with non-integer 
spatial dimensions.

The differential operators of the first order such as 
the gradient (\ref{grad-alpha}), 
the divergence (\ref{div-alpha}), 
the curl operator (\ref{curl-alpha}), and
the second order differential operators such as 
the scalar Laplacian (\ref{SLap-alpha}) and 
the vector Laplacian (\ref{VLap-alpha}),
allow us to describe 
anisotropic fractal media and materials
in the framework of continuum models 
with non-integer spatial dimensions.


\section{Examples of application}

\subsection{Poisson's equation}

Let us consider the Poisson's equation
for a fractal medium that is distributed along 
the positive half-$X$-axis
\be \label{Pois-1}
^S\Delta_{(\alpha)} \varphi (x) =f(x) .
\ee
Here we use the Laplace operator suggested in this paper.
Equation (\ref{Pois-1}) for single-variable case 
can be written as
\be
\frac{1}{c^2_1(\alpha,x)}
\left( \frac{\partial^2 \varphi(x)}{\partial x^2} 
- \frac{\alpha-1}{x} \varphi (x) \right) = f(x) 
\quad (x>0).
\ee
The general solution of this equation is
\be
\varphi (x) =C_1 + C_2 \, x^{\alpha} 
- \frac{\pi^{\alpha}}{\alpha (\Gamma(\alpha/2))^2}
\Bigl( 
\int f(x) \, x^{2 \alpha-1} \, dx -
x^{\alpha} \, \int f(x) \, x^{\alpha-1} \, dx 
\Bigr) .
\ee

Let us compare this result with the solution of 
Poisson's equation Laplace operator ${\cal K}_2$.
The Poisson's equation of the form
\be \label{Pois-1b}
{\cal K}_2 \varphi (x) =f(x) 
\ee
has the general solution 
\be
\varphi (x) =C_1 \, x^{(3-\alpha)/2} + C_2 \, x^{(1-\alpha)/2} 
+ x^{(3-\alpha)/2} \, \int f(x) \, x^{(1- \alpha)/2} \, dx -
x^{(1-\alpha)/2} \, \int f(x) \, x^{(1+\alpha)/2} \, dx .
\ee
In equation (\ref{Pois-1b}), 
we use the Laplace operator (\ref{NEW-S})
for single-variable case.

\subsection{Timoshenko beam equations for fractal material}

The Euler-Bernoulli beam theory gives 
a simplification of the linear theory of elasticity, 
which provides a means of calculating the load-carrying and deflection characteristics of beams and it covers the case for small deflections of a beam, which is subjected 
to lateral loads only.
Note that the Timoshenko beam equation for fractal materials
is discussed in the papers \cite{MOS-4,MOS-7}. 

In the Timoshenko beam theory without axial effects, 
the displacement vector ${\bf u}(x,y,z,t)$ 
of the beam are assumed to be given by
\be 
u_x(x,y,z,t) = -z \, \varphi(x,t) \, \quad 
u_y(x,y,z,t) = 0 , \quad 
u_z(x,y,t) = w(x,t) ,
\ee
where $(x,y,z)$ are the coordinates of a point in the beam, 
$u_x$, $u_y$, $u_z$ are the components of the displacement vector ${\bf u}$ , 
$\varphi=\varphi(x,t)$ is the angle of rotation of the normal to the mid-surface of the beam, 
and $w=w(x,t)$ is the displacement of the mid-surface in the $z$-direction.

Using a model with non-integer dimensional space 
for fractal media, we can use the derivatives 
\be \label{Der-New}
\partial_{x,\alpha} = c^{-1}_1(\alpha_x,x) \, D^{1}_x ,  
\quad \partial^n_{x,\alpha} =(\partial_{x,\alpha} )^n \quad 
(n \in \mathbb{N}) ,
\ee
where $c_1(\alpha_x,x)$ is defined for
non-integer dimensional space by
\be \label{c-1-Hb}
c_1(\alpha_x,x) = \frac{\pi^{\alpha_x/2}}{\Gamma (\alpha_x/2)} \, |x|^{\alpha_x-1} 
\ee
instead of the usual derivatives $D^1_x$ and $D^n_x$ 
for fractal materials.
If we use the derivatives (\ref{Der-New}), 
then the Timoshenko equation for fractal beam
can be derived  from the force and moment balance equations
\be \label{FMBE-2}
\rho \, A \, D^2_t w = \partial_{x,\alpha} Q , \quad 
\rho \, I^{(d)} \, D^2_t \varphi = Q - \partial_{x,\alpha}  M ,
\ee
with the bending moment
\be
M = - \, E\, I^{(d)} \, \partial_{x,\alpha} \varphi ,
\ee
and the shear force 
\be
Q = k \, G \, A \, \Bigr( \partial_{x,\alpha} w - \varphi \Bigr).
\ee
Here $I^{(d)}$ is the second moment of the fractal beam's cross-section.

The Timoshenko equations for homogeneous fractal beam 
has the form
\be \label{GTBE-1}
\rho \, A \, D^2_t w = k\, G\, A\, \partial_{x,\alpha} 
(\partial_{x,\alpha} w -\varphi ) ,
\ee
\be \label{GTBE-2}
\rho \, I^{(d)} \, D^2_t \varphi = 
k \, G\, A \, (\partial_{x,\alpha} w -\varphi ) + 
E \, I^{(d)} \, \partial^2_{x,\alpha} \varphi .
\ee

The Timoshenko fractal beam equations 
(\ref{GTBE-1}) and (\ref{GTBE-2}) also can be derived 
from the variational principle. 
The Lagrangian of the Timoshenko fractal beam has the form
\[ \mathcal{L}_{GTFB} 
= \frac{1}{2} \rho \, I^{(d)} \, \left( D^1_t \varphi(x,t) \right)^2 +
\frac{1}{2} \rho \, A \, \left( D^1_t w(x,t) \right)^2 -
\]
\be \label{Lagr-TB-2} 
- \frac{1}{2} (k G A) \, \left( \partial_{x,\alpha} w(x,t) - \varphi(x,t)  \right)^2 
 - \frac{1}{2} (EI^{(d)}) \, \left( \partial_{x,\alpha} \varphi(x,t) \right)^2  .
\ee
The stationary action principle gives the equations
\be \label{TB-2-5}
\frac{\partial\mathcal{L}}{\partial w} - 
D^1_t \, \left(\frac{\partial \mathcal{L}}{\partial D^1_t w}\right)  
- \, D^{1}_x \, \left(\frac{\partial \mathcal{L}}{\partial D^{1}_x w}\right) = 0 ,
\ee
\be \label{TB-2-6}
\frac{\partial\mathcal{L}}{\partial \varphi } - 
D^1_t \, \left(\frac{\partial \mathcal{L}}{\partial D^1_t \varphi }\right)  
- D^{1}_x \, \left(\frac{\partial \mathcal{L}}{\partial D^{1}_x \varphi }\right) = 0 .
\ee
Equations (\ref{TB-2-5} - \ref{TB-2-6}) are
the Euler-Lagrange equation for the 
model of fractal material described 
by the Lagrangian (\ref{Lagr-TB-2}).
Substitution of (\ref{Lagr-TB-2}) into
equations (\ref{TB-2-5} - \ref{TB-2-6}) gives
the Timoshenko fractal beam equations 
(\ref{GTBE-1}) and (\ref{GTBE-2}) that can be presented as
\be \label{GTBE-1b}
\rho \, A \, D^2_t w = k\, G\, A\, \partial_{x,\alpha} \, 
\Bigr( \partial_{x,\alpha} w - \varphi \Bigr) ,
\ee
\be \label{GTBE-2b}
\rho \, I^{(d)} \, D^2_t \varphi = 
k \, G\, A \, \Bigl(\partial_{x,\alpha} w -\varphi \Bigr) + 
E \, I^{(d)} \, \partial^2_{x,\alpha} \varphi .
\ee
If $\alpha=1$ then equations (\ref{GTBE-1b}-\ref{GTBE-2b}) are the Timoshenko equations for beam with homogeneous non-fractal material. 

For models with non-integer dimensional spaces, solutions of equations for fractal materials can be obtained from solutions of equations for non-fractal materials. Let $w_c(x,t)$ and $\varphi_c(x,t)$  be solutions of (\ref{GTBE-1b}-\ref{GTBE-2b}) with $\alpha=1$ and $x>0$, i.e., the Timoshenko equations for homogeneous non-fractal beam. Then solutions $w_F(x,t)$ and $\varphi_F(x,t)$  
of equations (\ref{GTBE-1b}-\ref{GTBE-2b}) for fractal beam with $0<\alpha <1$ can be represented by
\be
w_F(x,t)=w_c \left(\frac{\pi^{\alpha_x/2}}{\Gamma (\alpha_x/2)} \, |x|^{\alpha_x-1} ,t \right), \quad
\varphi_F(x,t)=\varphi_c
\left(\frac{\pi^{\alpha_x/2}}{\Gamma (\alpha_x/2)} \, |x|^{\alpha_x-1} 
,t \right) .  
\ee

\subsection{Euler-Bernoulli fractal beam}

As an example, we consider the equation for the Euler-Bernoulli homogeneous fractal beam
in the absence of a transverse load ($q(x)=0$), 
\be \label{107}
\rho \, A\, D^2_t w(x,t)+ E\, I^{(d)} \, \partial^4_{x,\alpha} w(x,t) = 0.
\ee 
This equation can be solved using the Fourier decomposition of 
the displacement into the sum of harmonic vibrations of the form
$w(x,t) = \text{Re}[w(x) \, \text{exp} (-i\omega t)]$,
where $\omega$ is the frequency of vibration. 
Then, equation (\ref{107}) gives 
the ordinary differential equation
\be
- \rho\, A\, \omega^2 w(x)
+E\, I^{(d)} \, \partial^4_{x,\alpha} w(x) = 0 .
\ee
The boundary conditions for fractal beam of length $L$ 
fixed at $x = 0$ are
\be \label{BC-1}
w(0)= 0 , \quad (\partial^1_{x,\alpha} w)(0) = 0 , 
\ee
\be \label{BC-2}
(\partial^2_{x,\alpha} w)(L) = 0 , \quad
(\partial^3_{x,\alpha} w)(L) = 0 .
\ee
The solution for the Euler-Bernoulli homogeneous fractal beam is defined by
\be \label{SOL-1}
w_{F,n}(x) = w_0 \, \left(\cosh (k_n x^{\alpha}) - \cos (k_n x^{\alpha}) +
C_n(\alpha) \ \Bigl( \sin (k_n x^{\alpha}) - \sinh (k_n x^{\alpha}) \Bigr) \right) , 
\quad x \in [0;L] ,
\ee
where $w_0$ is a constant, and
\be
C_n(\alpha)=\frac{\cos (k_n L^{\alpha}) + \cosh (k_n L^{\alpha})}{\sin (k_n L^{\alpha}) + \sinh (k_n L^{\alpha})} , 
\qquad
k_n = \frac{\pi^{\alpha/2}}{\Gamma(\alpha/2)} 
\, \left(\frac{\rho \, A\, \omega_n^2}{E\, I^{(d)}}\right)^{1/4} .
\ee
For boundary conditions (\ref{BC-1}-\ref{BC-2}),  
the solution (\ref{SOL-1}) exist only if $k_n$ are defined by
\be
\cosh(k_n L)\,\cos(k_n L) + 1 = 0 .
\ee
This trigonometric equation is solved numerically. 
The corresponding natural frequencies of vibration are
$\omega_n = k_n^2 \sqrt{(E \, I^{(d)}) /\rho \, A}$.
For a non-trivial value of the displacement, $w_0$ has to remain 
arbitrary, and the magnitude of the displacement is unknown for 
free vibrations. Usually $w_0=1$ is used when plotting mode 
shapes. 

\section{Conclusion}

We give a review of possible approaches to describe 
anisotropic fractal media.
We focused on two approaches based on the fractional space 
and the non-integer dimensional space.
There approaches allow us to describe anisotropic fractal 
media and materials in the framework of continuum models
by using the concept of density of states 
\cite{TarasovSpringer} and the product measure method.
Fractal medium is considered as a medium 
with non-integer mass dimension. 
The non-integer dimensionality
is a main characteristic property of fractal materials.
Therefore we suggest an application of differentiation 
and integration over non-integer dimensional spaces 
as natural way to describe fractal media 
\cite{CNSNS2014,CSF2014}.
Although, the non-integer dimension does not reflect 
all specific properties of real fractal materials, 
it allows us to formulate continuum models 
to derive important conclusions about the behavior of the media. 
In this paper a generalization of the
vector calculus for multi-fractional 
and non-integer dimensional spaces 
is proposed as tools to describe
anisotropic fractal media and materials  
in the framework of continuum models.
We suggest a generalization of vector calculus
for non-integer dimensional space that is product
of spaces with different dimensions.
The product measure method allows us to describe 
anisotropic fractal materials
by taking into account various non-integer 
dimensions in different directions.
The differential operators of the first order such as 
the gradient (\ref{grad-alpha}), 
the divergence (\ref{div-alpha}), 
the curl operator (\ref{curl-alpha}), and
the second order differential operators such as 
the scalar Laplacian (\ref{SLap-alpha}) and 
the vector Laplacian (\ref{VLap-alpha}),
are suggested to describe anisotropic fractal media and 
materials by continuum models 
with non-integer dimensional spaces.
To demonstrate some simple applications of
proposed approach to the description of 
fractal materials, we consider 
the Poisson's equation for fractal medium, 
the Euler-Bernoulli fractal beam and
the Timoshenko beam equations for fractal material.




\end{document}